\documentclass[12pt]{iopart}
\usepackage[dvips]{graphicx}
\usepackage{amssymb}
\usepackage{amsfonts}
\newcommand {\dr}{{\mathrm d}\mathbf{r}}

\newcommand {\dk}{{\mathrm d}\mathbf{k}}
\newcommand {\dpp}{{\mathrm d}\mathbf{p}}
\newcommand {\dd}{{\mathrm d}}
\newcommand {\rr}{\mathbf{r}}
\newcommand {\kk}{\mathbf{k}}
\newcommand {\qq}{\mathbf{q}}
\newcommand {\pp}{\mathbf{p}}
\newcommand {\X}{\mathbf{X}}
\newcommand {\F}{\mathbf{F}}
\newcommand {\jj}{\mathbf{j}}
\newcommand {\1}{\mathbf{1}}
\newcommand {\ee}{{\mathrm e}}

\begin{document}

\title{Dynamical density functional theory for dense atomic liquids}

\author{A.J. Archer\footnote{andrew.archer@bristol.ac.uk}}
\address{H.H. Wills Physics Laboratory, University of Bristol, Bristol
BS8 1TL, UK}
\date{\today}

\begin{abstract}
Starting from Newton's equations of motion,
we derive a dynamical density functional theory (DDFT) applicable to atomic
liquids. The theory has the feature that it requires as input
the Helmholtz free energy functional from equilibrium density functional theory.
This means that, given a reliable equilibrium free energy functional,
the correct equilibrium fluid density profile is guaranteed.
We show that when the isothermal compressibility is small,
the DDFT generates the correct value for the speed of sound in
a dense liquid. We also interpret the theory as a
dynamical equation for a coarse grained fluid density and show that the
theory can be used (making further approximations) to derive the standard
mode coupling theory that is used
to describe the glass transition. The present theory should provide a useful
starting point for describing the dynamics of inhomogeneous atomic fluids.
\end{abstract}


\section{Introduction}

Classical density functional theory (DFT) \cite{Evans79,Evans92}
allows one to determine the equilibrium ensemble average
one--body density profile of a fluid in the presence of an arbitrary
external field $V^{ext}(\rr)$. As DFT is based upon a minimisation
principle for a free energy density functional \cite{Evans79,Evans92},
one can use DFT to calculate
thermodynamic quantities such as surface tensions.
In practice, DFT is not exact because the exact
free energy functional is generally unknown. Nonetheless, the theory has proved
to be very powerful and there exist in the literature a large number of rather
accurate approximations for the Helmholtz free energy functional \cite{Evans92}
and equilibrium DFT is a well developed theoretical framework for describing
static fluid phenomena \cite{HM}. Given
this body of knowledge, constructing a dynamical density functional theory
(DDFT), that
requires as input the same Helmholtz free energy functional as the equilibrium
DFT, for the dynamics of the one--body density profile of an
inhomogeneous fluid that is out of equilibrium, is a very appealing idea.
In recent work \cite{MT,MT2}, Marconi and Tarazona derived an equation of motion
for the one body density profile of a fluid of particles with stochastic
equations of motion (i.e.~for colloidal fluids).
This theory has been applied to a number of problems
\cite{joe:christos, Archer7, Archer8, flor2, flor1,
Archer11, Archer13, RexetalPRE2005, RexetalMolP2006, Matthiasetal} and has
proved to be rather accurate where comparison with the results of Brownian
dynamics simulations has been made.
The question that then naturally arises is, can one construct a DDFT for a fluid
of particles with motion governed by Newton's equations? In a recent paper,
Chan and Finken \cite{ChanFinken} proved that the non-equilibrium one body
density $\rho(\rr,t)$ is uniquely determined for a given time dependent external
field $V^{ext}(\rr,t)$, where $t$ is the time.
They showed that in general the DDFT requires a
non--equilibrium free energy functional that is not the same as the free energy
functional entering equilibrium DFT and that the (approximate) theory of
Marconi and Tarazona \cite{MT,MT2} can be obtained by taking the
adiabatic limit of the `exact' DDFT \cite{ChanFinken}.
Given that very little is known about the non--equilibrium free energy
functional, this makes practical calculations of the
fluid dynamics in the Chan--Finken framework somewhat difficult.

In the present paper, we derive an approximate DDFT for a fluid of particles
described by Newton's equations of motion (i.e.~an atomic fluid, as opposed to a
colloidal fluid). Our DDFT requires as input the equilibrium
free energy functional and therefore makes practical calculation of the fluid
dynamics much more tractable. The theory is an approximate one,
and in the form derived here is most relevant to dense fluids.
On applying the theory, in the limit that the ratio of the heat capacities
$C_P/C_V \rightarrow 1$ (i.e.~when the isothermal compressibility is small),
we find that the DDFT reproduces the correct value for the speed of sound
waves in a dense liquid, and predicts the length scale over which
sound waves are attenuated. We also interpret the theory as one for a
coarse grained one--body density profile, $\bar{\rho}(\rr,t)$, and from these
equations we derive (making a number of further approximations) the mode
coupling theory (MCT) that is used to describe the glass transition
\cite{HM,Gotze}. We believe that the present theory will provide a useful
starting point for determining the dynamics of the one body density profile
of an atomic fluid.

This paper is laid out as follows: In Sec.~\ref{sec:2}, starting from Newton's
equations of motion, we derive equations of motion for the ensemble average
fluid one body density profile $\rho(\rr,t)$. In Sec.~\ref{sec:3}, we introduce
a number of approximations in order to close the equations derived in
Sec.~\ref{sec:2}. This closure
yields us the DDFT. In Sec.~\ref{sec:4} we apply the DDFT
in two particular situations: (i) sound waves propagating in a fluid and (ii) to
dynamic two point correlation functions, where we derive the MCT.
Finally, in Sec.~\ref{sec:conc} we draw some conclusions.

\section{Density equations of motion}
\label{sec:2}

The (classical) density operator for a system of $N$ identical particles is
$\hat{\rho}(\rr,t)=\sum_i\delta(\rr-\rr_i(t))$, where $\rr^N=\{\rr_i(t)\}$ is
the set of position coordinates of the particles and the index $i$ labels the
different particles. In Fourier space, the density
operator is given by:
\begin{equation}
\rho_\kk(t)=\int \dr \ee^{i\kk\cdot \rr}\hat{\rho}(\rr,t)
=\sum_i \ee^{i\kk\cdot \rr_i}.
\label{eq:rho_k}
\end{equation}
Newton's equations of motion for these particles are
\begin{equation}
\ddot{\rr}_i(t)=-\frac{1}{m}\frac{\partial V(\rr^N,t)}{\partial \rr_i},
\label{eq:Newton}
\end{equation}
where $m$ is the mass of a particle and
\begin{equation}
V(\rr^N,t)=\sum_iV^{ext}(\rr_i,t)+\frac{1}{2}\sum_{i,j}v(\rr_i(t)-\rr_j(t))
\label{eq:PE}
\end{equation}
is the potential energy of the system. We assume this is composed of one body
contributions
due to the external potential $V^{ext}(\rr,t)$ and a sum of pair interactions
between the particles; $v(r)$ is the pair potential. Following Zaccarelli
\etal~\cite{Zaccarelli}, we take two derivatives of Eq.~(\ref{eq:rho_k}) to
obtain:
\begin{equation}
\ddot{\rho}_\kk(t)=
-\sum_i (\kk \cdot \dot{\rr}_i(t))^2\ee^{i\kk\cdot \rr_i}
+\sum_i i(\kk \cdot \ddot{\rr}_i(t))\ee^{i\kk\cdot \rr_i}.
\label{eq:rho_k_ddot}
\end{equation}
We can express the potential energy as:
\begin{equation}
V(\rr^N,t)=\sum_i \int \frac{\dk}{(2 \pi)^3}V^{ext}_\kk(t)\ee^{-i\kk\cdot \rr_i}
+\frac{1}{2}\sum_{i,j} \int \frac{\dk}{(2 \pi)^3}v_\kk\ee^{-i\kk\cdot
(\rr_i-\rr_j)},
\end{equation}
in an obvious notation. Differentiation gives:
\begin{equation}
\fl
\frac{\partial V(\rr^N,t)}{\partial \rr_i}
=\int \frac{\dk}{(2 \pi)^3}V^{ext}_\kk(t)(-i\kk \ee^{-i\kk\cdot \rr_i})
+\sum_{j} \int \frac{\dk}{(2 \pi)^3}v_\kk(-i\kk \ee^{-i\kk\cdot
(\rr_i-\rr_j)}).
\label{eq:grad_V}
\end{equation}
Combining Eqs.~(\ref{eq:Newton}), (\ref{eq:rho_k_ddot}) and (\ref{eq:grad_V}) we
obtain the following equation \cite{Zaccarelli}:
\begin{eqnarray}
\fl
\ddot{\rho}_\kk(t)=
-\sum_i (\kk \cdot \dot{\rr}_i(t))^2\ee^{i\kk\cdot \rr_i}
-\frac{1}{m}\sum_{i,j} \int \frac{\dk'}{(2 \pi)^3}v_{\kk'}(\kk\cdot\kk')
\ee^{-i\kk' \cdot(\rr_i-\rr_j)}\ee^{i\kk \cdot \rr_i} \nonumber \\
-\frac{1}{m}\sum_i\int \frac{\dk'}{(2 \pi)^3}V^{ext}_{\kk'}(t)(\kk\cdot\kk')
\ee^{-i\kk'\cdot \rr_i}\ee^{i\kk \cdot \rr_i}.
\label{eq:Zacc_1}
\end{eqnarray}
Using Eq.~(\ref{eq:rho_k}) this can be rewritten as:
\begin{eqnarray}
\fl
\ddot{\rho}_\kk(t)=
-\sum_i (\kk \cdot \dot{\rr}_i(t))^2\ee^{i\kk\cdot \rr_i}
-\frac{1}{m}\int \frac{\dk'}{(2 \pi)^3}v_{\kk'}(\kk\cdot\kk')
\rho_{\kk-\kk'}(t)\rho_{\kk'}(t) \nonumber \\
-\frac{1}{m}\int \frac{\dk'}{(2 \pi)^3}V^{ext}_{\kk'}(t)(\kk\cdot\kk')
\rho_{\kk-\kk'}(t).
\label{eq:Zacc}
\end{eqnarray}
This equation is simply a generalisation to the case with an external field of
Eq.~(8) in Ref.~\cite{Zaccarelli}. We transform Eq.~(\ref{eq:Zacc}) back into
real space and obtain the following:
\begin{eqnarray}
\fl
\ddot{\hat{\rho}}(\rr,t)=
-\sum_i\int \frac{\dk}{(2 \pi)^3}
(\kk \cdot \dot{\rr}_i(t))^2\ee^{i\kk\cdot (\rr_i-\rr)}
+\frac{1}{m}\nabla \cdot \int \dr' \hat{\rho}(\rr,t)
\hat{\rho}(\rr',t)\nabla v(\rr-\rr') \nonumber \\
+\frac{1}{m} \nabla \cdot \left[\hat{\rho}(\rr,t) \nabla V^{ext}(\rr,t)\right].
\label{eq:Zacc_a}
\end{eqnarray}
Our interest is in ensemble average quantities. We therefore average in
Eq.~(\ref{eq:Zacc_a}) over the ensemble of initial momenta and positions to
obtain an equation of motion for the ensemble average one body density,
$\rho(\rr,t)=\langle \hat{\rho}(\rr,t) \rangle$:
\begin{eqnarray}
\fl
\ddot{\rho}(\rr,t)=
-\langle \sum_i\int \frac{\dk}{(2 \pi)^3}
(\kk \cdot \dot{\rr}_i(t))^2\ee^{i\kk\cdot (\rr_i-\rr)}\rangle
+\frac{1}{m}\nabla \cdot \int \dr' 
\rho^{(2)}(\rr,\rr',t)\nabla v(\rr-\rr') \nonumber \\
+\frac{1}{m} \nabla \cdot \left[\rho(\rr,t) \nabla V^{ext}(\rr,t)\right],
\label{eq:av_Zacc}
\end{eqnarray}
where $\langle \cdots \rangle$ denotes the ensemble average and
where $\rho^{(2)}(\rr,\rr',t)=\langle \hat{\rho}(\rr,t)\hat{\rho}(\rr',t)
\rangle$ is the ensemble average, time dependent, two body distribution
function. We now consider the first term on the right hand side of
Eq.~(\ref{eq:av_Zacc}). This contains the expression
$\kk \cdot \dot{\rr}_i(t)=k \dot{r}_{i,par}(t)$, where $\dot{r}_{i,par}(t)$
is the component of the velocity of the $i^{{\mathrm th}}$ particle in the
direction of $\kk$. We write $\dot{r}_{i,par}^2(t)=\langle
\dot{r}_{i,par}^2\rangle+\alpha_i(t)$, where
$\alpha_i(t)$ is just the fluctuation of $\dot{r}_{i,par}^2(t)$ about its mean
equilibrium value. Note also that the equilibrium quantity
$\langle \dot{r}_{i,par}^2\rangle=k_BT/m$, from
equipartition. We may therefore re-write the first term on the right hand side
of Eq.~(\ref{eq:av_Zacc}) as follows:
\begin{eqnarray}
\fl
\langle \sum_i\int \frac{\dk}{(2 \pi)^3}
(\kk \cdot \dot{\rr}_i(t))^2\ee^{i\kk\cdot (\rr_i-\rr)}\rangle
=\int \frac{\dk}{(2 \pi)^3} k^2
\langle \sum_i \alpha_i(t) \ee^{i\kk\cdot \rr_i}\rangle \ee^{-i\kk\cdot \rr}
\nonumber \\
+\frac{k_BT}{m}\int \frac{\dk}{(2 \pi)^3} k^2
\langle \sum_i \ee^{i\kk\cdot \rr_i}\rangle \ee^{-i\kk\cdot \rr} \nonumber \\
=B(\rr,t)-\frac{k_BT}{m}\nabla^2\rho(\rr,t).
\label{eq:aa}
\end{eqnarray} 
In order to obtain the second line we have used Eq.~(\ref{eq:rho_k}) and defined
\begin{eqnarray}
B(\rr,t) \equiv \int \frac{\dk}{(2 \pi)^3} k^2
\langle \sum_i \alpha_i(t) \ee^{i\kk\cdot \rr_i}\rangle \ee^{-i\kk\cdot \rr}.
\label{eq:B}
\end{eqnarray}
Substituting Eq.~(\ref{eq:aa}) into Eq.~(\ref{eq:av_Zacc}) we obtain
\begin{eqnarray}
\fl
\ddot{\rho}(\rr,t)+B(\rr,t)=
\frac{k_BT}{m}\nabla^2\rho(\rr,t)
+\frac{1}{m}\nabla \cdot \int \dr' 
\rho^{(2)}(\rr,\rr',t)\nabla v(\rr-\rr')\nonumber \\
+ \frac{1}{m}\nabla \cdot \left[\rho(\rr,t) \nabla V^{ext}(\rr,t)\right].
\label{eq:DDFT_1}
\end{eqnarray}
This equation, taken together with Eq.~(\ref{eq:B}) is still formally exact. We
note that the right hand side of this equation is simply
the derivative of the first equation of the YBG hierarchy \cite{HM}.
This means that at equilibrium the right hand side of Eq.~(\ref{eq:DDFT_1}) is
equal to zero and therefore at equilibrium $B(\rr,t)=0$.
In the Appendix we present an alternative derivation of Eq.~(\ref{eq:DDFT_1}),
starting from the first equation of the BBGKY hierarchy \cite{HM}.
The derivation gives us an alternative expression for
$B(\rr,t)$, that may provide further insight into the nature of this
term.

\section{The DDFT}
\label{sec:3}

Eq.~(\ref{eq:DDFT_1}) is the starting point for our derivation of the DDFT.
We make two approximations in Eq.~(\ref{eq:DDFT_1}) in order to render
a closed and tractable DDFT. Our first assumption is the same as
that used to derive the the colloidal DDFT \cite{MT,MT2,Archer7,Archer8}:
we assume that the
two particle correlations in the non-equilibrium fluid are the same as
in an equilibrium fluid with the same one body density profile, i.e.~we assume
that we can apply the following sum rule, which is exact for an equilibrium
fluid \cite{Archer7}: 
\begin{eqnarray}
- \, k_BT \rho(\rr) \nabla c^{(1)}(\rr)
= \int \dr' \rho^{(2)}(\rr,\rr') \nabla v(\rr,\rr'),
\label{eq:grad_c1}
\end{eqnarray}
where $c^{(1)}(\rr)$ is the one body direct correlation function and is
equal to the functional derivative of the excess (over ideal) part of the
Helmholtz free energy functional \cite{Evans79,Evans92}:
\begin{equation}
c^{(1)}(\rr) \, = \, -\beta
\frac{\delta F_{ex}[\rho(\rr)]}{\delta \rho(\rr)}.
\label{eq:c1}
\end{equation}
We must also make an approximation for the term $B(\rr,t)$ in
Eq.~(\ref{eq:DDFT_1}). Since at equilibrium $B(\rr,t)=0$, the simplest
approximation we can make is to assume $B(\rr,t) \propto
\dot{\rho}(\rr,t)$. Making these two approximations we obtain our main
result:
\begin{equation}
\frac{\partial^2 \rho(\rr,t)}{\partial t^2}
+\nu \frac{\partial \rho(\rr,t)}{\partial t}=
\frac{1}{m}\nabla \cdot \left[ \rho(\rr,t) \nabla \frac{\delta
F[\rho(\rr,t)]}{\delta \rho(\rr,t)} \right],
\label{eq:mainres}
\end{equation}
where $\nu$ is an undetermined (collision) frequency.
In general, we expect $\nu$ to be a (non-local) functional of
the density $\rho(\rr,t)$. We show below that when density modulations are
small, then the collision frequency $\nu=k_BT/mD$, where $D$ is the
self diffusion coefficient. $F$ is the Helmholtz free energy functional:
\begin{eqnarray}
F[{\rho}(\rr,t)]=F_{id}[{\rho}(\rr,t)]
+ F_{ex}[{\rho}(\rr,t)] + \int \dr V_{ext}(\rr,t) {\rho}(\rr,t),
\label{eq:F}
\end{eqnarray}
where
\begin{eqnarray}
F_{id}[{\rho}(\rr,t)]
= k_BT \int \dr {\rho}(\rr,t)[\ln({\rho}(\rr,t) \Lambda^3)-1]
\label{eq:F_id}
\end{eqnarray}
is the ideal gas contribution to the free energy, and
where $\Lambda$ is the thermal
de Broglie wavelength. We must emphasise that strictly speaking, $F$ is defined
as a functional of the equilibrium ensemble
average density density profile $\rho(\rr)$. We are effectively
assuming that we may use $F$ as an approximation for the unknown dynamical
free energy functional. It is important to note that at equilibrium, given a
reliable approximation for $F$, Eq.~(\ref{eq:mainres}) guarantees the exact
equilibrium fluid density profile \cite{Archer7}. For the case when the
potential energy $V(\rr^N,t)$ also contains three--body and higher body
interactions, as we show in the Appendix, one still obtains
Eq.~(\ref{eq:mainres}), making approximations equivalent to those made above.
The argument used is somewhat similar to the approach used in
Ref.~\cite{Archer7} to generalise the DDFT of Marconi and Tarazona
\cite{MT,MT2} to the case with multi--body interactions.

In order to show that when density modulations are small the coefficient
$\nu=k_BT/mD$, we consider a fluid in which a (small) number, $n$,
of the particles are labelled in such a way that we can think of them as a
different species of particles, without changing
any of the interaction potentials between the particles. If at time $t=0$ the
$n$ labelled particles are confined to a small volume near the origin $(\rr=0)$,
then in the limit $t\rightarrow \infty$ the density profile of these particles
will be $\rho_l(r,t)=nG_s(r,t)$, where $l$ denotes the labelled particles and
$G_s(r,t)$ is the van Hove self correlation function \cite{HM}. In this limit
the one body density of the remaining particles $\rho(r,t) \simeq \rho_b$, where
$\rho_b$ is the bulk fluid density. It is known that in the limit $t \rightarrow
\infty$ that \cite{HM}
\begin{equation}
G_s(r,t)=\frac{1}{(4 \pi D t)^{3/2}} \exp \left(\frac{-r^2}{4 Dt}\right).
\label{eq:G_s}
\end{equation}
Using this result, together with Eq.~(\ref{eq:mainres}) generalised to a two
component fluid, we obtain the following after making a Taylor expansion of the
excess Helmholtz free energy functional:
\begin{equation}
\frac{\partial^2 \rho_l(\rr,t)}{\partial t^2}
+\nu \frac{\partial \rho_l(\rr,t)}{\partial t}=
\frac{k_BT}{m}\nabla^2 \rho_l(\rr,t) + \cdots,
\label{eq:taylor_mainres}
\end{equation}
where $\cdots$ denotes terms of order $\rho_l^2$, $\rho_l(\rho-\rho_b)$ and
higher order terms. Furthermore, in the limit $t \rightarrow \infty$,
$\dot{\rho_l} \gg \ddot{\rho_l}$, so we may neglect the $\ddot{\rho_l}$ term.
Substituting Eq.~(\ref{eq:G_s}) into
Eq.~(\ref{eq:taylor_mainres}) we obtain the result that $\nu=k_BT/mD$.

At this point, we note that in cases where the collision frequency
$\nu$ is large, then $\nu \dot{\rho}
\gg \ddot{\rho}$ and we may, in certain circumstances, approximate
Eq.~(\ref{eq:mainres}) by
\begin{equation}
\nu \frac{\partial \rho(\rr,t)}{\partial t}=
\frac{1}{m}\nabla \cdot \left[ \rho(\rr,t) \nabla \frac{\delta
F[\rho(\rr,t)]}{\delta \rho(\rr,t)} \right].
\label{eq:mainres_colloid}
\end{equation}
This is equivalent to the DDFT of Marconi and Tarazona \cite{MT,MT2},
which was derived for colloidal fluids (fluids of
Brownian particles) -- see also the discussion in Ref.~\cite{Yoshimori}.

\section{Applications of the DDFT}
\label{sec:4}

Having argued for considering Eq.~(\ref{eq:mainres}) as an approximation to the
equation governing the dynamics of the particles, we now examine some of the
consequences. The first case we consider is that of sound waves
propagating in a dense fluid.

\subsection{Sound waves in a dense liquid}

We Taylor expand the excess
Helmholtz free energy functional in terms of density modulations $\delta
\rho(\rr,t)$ about a uniform bulk density $\rho_b$ (i.e.~we write
$\rho(\rr,t)=\rho_b+\delta \rho(\rr,t)$). Omitting terms beyond
${\cal O}(\delta \rho^2)$ we obtain \cite{Evans92,Archer7}:
\begin{eqnarray}
\fl
F_{ex}[\rho(\rr,t)] \,=\, F_{ex}[\rho_b] -c^{(1)}(\infty)
\int \dr \delta\rho(\rr,t) \nonumber \\
- \frac{k_BT}{2} \int \dr \int \dr'
\delta{\rho}(\rr,t)\delta{\rho}(\rr',t) c^{(2)}(\rr-\rr'),
\label{eq:F_ex}
\end{eqnarray}
where $c^{(2)}(r)$ is the Ornstein-Zernike pair direct correlation function.
Combining Eq.~(\ref{eq:F_ex}) with Eq.~(\ref{eq:mainres}) we obtain
\cite{Archer7}:
\begin{eqnarray}
\fl
\frac{\partial^2 \delta \rho(\rr,t)}{\partial t^2}
+\nu \frac{\partial \delta\rho(\rr,t)}{\partial t} =
\frac{k_BT}{m}\nabla^2 \delta \rho(\rr,t)
- \frac{k_BT\rho_b}{m} \nabla^2
\int \dr' \delta \rho(\rr',t) \nabla c^{(2)}(\rr-\rr').
\label{eq:abc}
\end{eqnarray}
Sound waves correspond to the following fluid one body density profile:
\begin{eqnarray}
\rho(\rr,t)&=&\rho_b+\delta \rho(\rr,t) \nonumber \\
&=&\rho_b+\epsilon \exp[i(\qq\cdot \rr-\omega t)],
\label{eq:rho_waves}
\end{eqnarray}
where $\epsilon$ is the amplitude, which is assumed to be
small compared to the bulk fluid
density $\rho_b$ and $|\qq|=k+i\lambda^{-1}$. The speed of sound in the liquid,
$c_s=\omega/k$, and $\lambda$ is the distance over which the amplitude of the
sound is attenuated by a factor $1/\ee$.
Substituting Eq.~(\ref{eq:rho_waves}) into Eq.~(\ref{eq:abc}) we obtain:
\begin{eqnarray}
\left(-\omega^2-i \omega \nu \right) \delta \rho
=\left(-\frac{q^2k_BT}{m}+\frac{q^2\rho_bk_BT \hat{c}(q)}{m}\right) \delta \rho,
\label{eq:abcd}
\end{eqnarray}
where $\hat{c}(q)$ is the Fourier transform of $c^{(2)}(r)$.
Equating the real and imaginary
parts on the left and right hand sides of Eq.~(\ref{eq:abcd}) we obtain the
dispersion relation:
\begin{eqnarray}
\omega^2(k)=\frac{k^2k_BT}{mS(k)},
\label{eq:dispersion}
\end{eqnarray}
and the following expression for the attenuation length:
\begin{eqnarray}
\lambda(k)=\frac{2\omega}{\nu k},
\label{eq:attenuation}
\end{eqnarray}
where we have used the fact that the fluid static structure factor
$S(k)=(1-\rho_b\hat{c}(k))^{-1}$ and we have assumed that $k\gg \lambda^{-1}$.

Sound waves correspond to the long wavelength limit. In this
limit, $S(k) \simeq S(0) = \rho_b k_BT \chi_T$,
i.e.~proportional to the isothermal compressibility $\chi_T$,
yielding
\begin{eqnarray}
c_s^2=\frac{1}{\rho_b m \chi_T}.
\label{eq:speed}
\end{eqnarray}
The adiabatic speed of sound in liquids is known to be \cite{HM}
\begin{eqnarray}
c_s^2=\frac{\gamma}{\rho_b m \chi_T},
\label{eq:speed_proper}
\end{eqnarray}
where $\gamma=C_P/C_V$, is the ratio of the heat capacities. The reason that our
expression (\ref{eq:speed}) for $c_s$ differs from the true value, given by
Eq.~(\ref{eq:speed_proper}), can be traced back to the approximation
$B(\rr,t) \simeq \nu\dot{\rho}(\rr,t)$, used to obtain
Eq.~(\ref{eq:mainres}) from Eq.~(\ref{eq:DDFT_1}). In making this approximation
we are effectively neglecting temperature fluctuations. However, 
the density fluctuations corresponding to sound waves occur on time scales too
fast to reach isothermal conditions; sound wave fluctuations are an
adiabatic processes. Using the fact that $C_P=C_V+T\chi_T \beta_V^2/\rho_b$
\cite{HM}, where $\beta_V$ is the thermal pressure coefficient,
we see that it is only
at low temperatures and high densities, when $\chi_T$ is small,
that $C_P \simeq C_V$, and Eq.~(\ref{eq:speed}) yields the correct value for
$c_s$, in agreement with Eq.~(\ref{eq:speed_proper}).

\subsection{DDFT and MCT}

Using DDFT (in various formulations) to describe the slow dynamics of fluids at
state points near to the glass transition has been suggested by Kawasaki and
coworkers \cite{KawasakiPhysicaA1994, KawasakiJPCM2000, KawasakiandKim,
FuchizakiKawasakiJPCM2002, KawasakiJStatP2003}.
In this section we show how the standard mode coupling theory (MCT)
can be obtained from a temporally coarse grained DDFT,
as opposed to the spatially coarse grained DDFT, described
in Ref.~\cite{KawasakiPhysicaA1994}. Our approach, in principle,
allows us to include the effects of external potentials in the MCT.
The arguments in this section are to some extent equivalent to those made in
Ref.~\cite{Zaccarelli}.

Eq.~(\ref{eq:mainres}), as derived above, is an equation for the ensemble
average fluid
one body density profile, $\rho(\rr,t)$. However, we may also interpret it as
an equation for the time evolution of a coarse grained density profile,
$\bar{\rho}(\rr,t)$. For example, we define a temporally coarse grained
density \cite{Archer8}:
\begin{equation}
\bar{\rho}(\rr,t)=\int_{-\infty}^\infty {\mathrm d}t' K(t-t')
\hat{\rho}(\rr,t'),
\label{eq:rho_bar}
\end{equation}
where $K(t)$ is a normalised function of finite width in time. Multiplying
Eq.~(\ref{eq:Zacc_a}) by $K(t-t')$ and then integrating over $t$,
we obtain an equation of the same form as
Eq.~(\ref{eq:av_Zacc}) for the coarse grained density $\bar{\rho}(\rr,t)$. This
equation involves the coarse grained two body density distribution function:
\begin{equation}
\bar{\rho}^{(2)}(\rr,\rr',t)=\int_{-\infty}^\infty {\mathrm d}t' K(t-t')
\hat{\rho}(\rr,t')\hat{\rho}(\rr',t').
\label{eq:rho2_bar}
\end{equation}
If we assume further that we have coarse grained sufficiently that we can
approximate the term involving $\bar{\rho}^{(2)}(\rr,\rr',t)$ using the sum
rule (\ref{eq:grad_c1}), and that the coarse grained equivalent of $B(\rr,t)$,
$\bar{B} \simeq \nu \bar{\rho}$, then we obtain Eq.~(\ref{eq:mainres}) as our
equation of motion for the coarse grained density $\bar{\rho}(\rr,t)$. We are
effectively assuming that we can approximate the 
coarse grained free energy functional by the Helmholtz free energy functional
entering equilibrium DFT -- i.e.~the free energy $F$ in Eq.~(\ref{eq:F}) is now
a functional of the coarse grained density $\bar{\rho}(\rr,t)$.

We now proceed to obtain the MCT, using a particular approximation for $F$,
namely the Taylor expansion in Eq.~(\ref{eq:F_ex}).
We will also assume for the time being that the external potential $V_{ext}=0$.
What follows is to some extent the argument of Kawasaki in Ref.~\cite{Kawasaki}.
Using Eq.~(\ref{eq:F_ex}) in Eq.~(\ref{eq:mainres}) we find \cite{Archer7}:
\begin{eqnarray}
\frac{\partial^2 \bar{\rho}(\rr,t)}{\partial t^2} 
+\nu \frac{\partial \bar{\rho}(\rr,t)}{\partial t}
-\frac{k_BT}{m}\nabla^2 \bar{\rho}(\rr,t) \nonumber \\
+\frac{k_BT}{m}\nabla \cdot \bigg[ (\rho_b +
\delta \bar{\rho}(\rr,t)) \int \dr' \delta \bar{\rho}(\rr',t) \nabla
c^{(2)}(|\rr-\rr'|;\rho_b)
\bigg]=0.
\label{eq:abcde}
\end{eqnarray}
Fourier transforming Eq.\ (\ref{eq:abcde}) we obtain \cite{Archer7}:
\begin{eqnarray}
\fl
\ddot{\rho}_{\kk}(t)+\nu \dot{\rho}_{\kk}(t)
=-\frac{k^2}{\beta m} \rho_{\kk}(t)+\frac{\rho_b k^2}{\beta m}
c_{\kk} \rho_{\kk}(t)
+\frac{1}{\beta m}\frac{1}{(2 \pi)^3} \int \dk' \kk\cdot \kk'
\rho_{\kk'}(t)c_{\kk'}\rho_{\kk-\kk'}(t),
\label{eq:Spin_dec_eq_2_next}
\end{eqnarray}
where $c_{\kk}$ is the Fourier transform of the pair direct correlation
function $c^{(2)}(r)$, obtained by taking two functional derivatives of
$F[\bar{\rho}(\rr,t)]$, and $\rho_{\kk}(t)$ denotes the Fourier transform of
$\delta \bar{\rho}(\rr,t)$. We rewrite Eq.~(\ref{eq:Spin_dec_eq_2_next}),
in the form:
\begin{eqnarray}
\ddot{\rho}_{\kk}(t)+\nu\dot{\rho}_{\kk}(t)
+\Omega_{\kk}^2 \rho_{\kk}(t)=\hat{R}_{\kk}(t),
\label{eq:sugg}
\end{eqnarray}
where $\Omega_{\kk}^2=k^2/\beta m S_{\kk}$ (we have defined the
static structure factor $S_{\kk}=[1-\rho_b c_{\kk}]^{-1}$) and where we have
defined
\begin{eqnarray}
\hat{R}_{\kk}(t)=\frac{1}{\beta m (2 \pi)^3} \int \dk' \kk \cdot \kk'
\rho_{\kk'}(t)c_{\kk'}\rho_{\kk-\kk'}(t).
\label{eq:R_hat}
\end{eqnarray}
Eq.~(\ref{eq:sugg}) is suggestive,
since it is known that one can always write formally
the equation of motion for a dynamical variable such as $\rho_{\kk}(t)$ in
the following form \cite{Kawasaki,Zwanzig}:
\begin{eqnarray}
\ddot{\rho}_{\kk}(t)+\nu\dot{\rho}_{\kk}(t)
+\Omega_{\kk}^2 \rho_{\kk}(t)
=-\int_0^t \dd t' m_{\kk}(t')\dot{\rho}_{\kk}(t-t')
+R_{\kk}(t),
\label{eq:dyn_eq}
\end{eqnarray}
where $R_{\kk}(t)$ is a noise term and the memory function is given formally by:
\begin{eqnarray}
m_{\kk}(t)=\frac{\langle R_{\kk}(t)R_{-\kk}(0)\rangle}
{\langle |\dot{\rho}_{\kk}(t)|^2\rangle},
\label{eq:memory_func}
\end{eqnarray}
where $\langle \cdots \rangle$ denotes the average over the ensemble of initial
configurations of the density. Since
$\dot{\rho}_{\kk}(t)=i \kk\cdot j_{\kk}(t)=ikj_\kk^L(t)$, where $j_\kk^L(t)$ is
the longitudinal current, the denominator in (\ref{eq:memory_func}) is simply
equal to $k^2Nk_BT/m$. One finds that $\rho_{-\kk}(0)$ and $R_{\kk}(t)$ are
statistically independent \cite{Kawasaki},
so on multiplying Eq.~(\ref{eq:dyn_eq}) by $\rho_{-\kk}(0)$, and then averaging,
one obtains the following dynamical equation:
\begin{eqnarray}
\ddot{\phi}_{\kk}(t)+\nu\dot{\phi}_{\kk}(t)
+\Omega_{\kk}^2 \phi_{\kk}(t)
=-\int_0^t \dd t' m_{\kk}(t')\dot{\phi}_{\kk}(t-t'),
\label{eq:MCT_eq}
\end{eqnarray}
where the normalised correlator
\begin{eqnarray}
\phi_{\kk}(t)=\frac{\langle \rho_{\kk}(t)\rho_{-\kk}(0)\rangle}
{\langle \rho_{\kk}(0)\rho_{-\kk}(0)\rangle}.
\label{eq:correlator}
\end{eqnarray}
Eq.~(\ref{eq:MCT_eq}) is formally exact. However, one does not have a tractable
form for the memory function $m_{\kk}(t')$.
It is at this point that we return to the DDFT Eqs.~(\ref{eq:sugg}) and
(\ref{eq:R_hat}) and we make the following {\em assumption}
\cite{Kawasaki}:
\begin{eqnarray}
\langle R_{\kk}(t)R_{-\kk}(0)\rangle
\simeq \langle \hat{R}_{\kk}(t)\hat{R}_{-\kk}(0)\rangle,
\label{eq:assumption}
\end{eqnarray}
i.e.~we will assume that the time correlations in $\hat{R}_{\kk}(t)$ and
$R_{\kk}(t)$ are the same \cite{Kawasaki}. We also assume the usual MCT
factorisation of four--point correlation functions into products of two--point
correlation functions \cite{HM,Gotze}:
$\langle \rho_{\kk'}(t) \rho_{\kk-\kk'}(t) \rho_{\kk''}(0)
\rho_{-\kk-\kk'}(0)\rangle \simeq \langle \rho_{\kk'}(t) \rho_{\kk''}(0)\rangle
\langle \rho_{\kk-\kk'}(t)\rho_{-\kk-\kk'}(0)\rangle
+\langle \rho_{\kk'}(t)\rho_{-\kk-\kk'}(0)\rangle
\langle\rho_{\kk-\kk'}(t) \rho_{\kk''}(0)\rangle$ and we obtain the standard MCT
expression for the memory function \cite{Gotze,Kawasaki}:
\begin{eqnarray}
\fl
m_{\kk}(t)=\frac{k_BT \rho_b}{2 (2 \pi)^3k^2m}
\int \dk'\left[\kk\cdot\kk'c_{\kk'}+\kk \cdot(\kk-\kk')c_{\kk-\kk'}\right]^2
S_{\kk'}S_{\kk-\kk'}\phi_{\kk'}(t)\phi_{\kk-\kk'}(t).
\label{eq:m}
\end{eqnarray}
In this analysis we have derived equations for determining the correlator
(\ref{eq:correlator}), where $\rho_{\kk}(t)$ is defined as the Fourier
transform of the coarse grained density. However, the standard MCT is a theory
for the correlator (\ref{eq:correlator}), with $\rho_{\kk}(t)$ given by
Eq.~(\ref{eq:rho_k}). Strictly speaking, we have derived a theory with exactly
the same structure, but for
a different correlator. However, given that much of the physics of dense fluids
near to the glass transition concerns the behaviour of collective density
fluctuations, it is not too surprising that a description in terms of a coarse
grained density is not that different from a description in terms
of the density field itself.

Returning to the
case when there is an external field, we find there is the following additional
term on the right hand side of Eq.~(\ref{eq:sugg}):
\begin{eqnarray}
E_{\kk}(t)=i \kk \cdot \int \dr \exp(i \kk \cdot \rr) \bar{\rho}(\rr,t)
\nabla V_{ext}(\rr,t).
\label{eq:extpot_term}
\end{eqnarray}
This results in additional terms in Eq.~(\ref{eq:dyn_eq}) as well as
modifying the memory function $m_{\kk}(t)$.
Consider an external potential $V_{ext}(\rr) \sim \exp(i \qq \cdot\rr)$,
i.e.~a periodic field such as that generated by a laser with wavelength
$2 \pi/q$. In this case we find
\begin{eqnarray}
E_{\kk}(t)\simeq- \epsilon\kk \cdot \qq \rho_{\kk+\qq}(t),
\label{eq:extpot_term_lazer}
\end{eqnarray}
where $\epsilon$ is proportional to the amplitude of the external field. This
term in Eq.~(\ref{eq:sugg}) couples density fluctuations with wavelength $\kk$
to fluctuations with wavelength $\kk+\qq$. Note also that the coupling is
strongest when $\kk$ is parallel to $\qq$.
Including such a contribution in Eq.~(\ref{eq:dyn_eq}), multiplying through by
$\rho_{-\kk}(0)$ and then averaging would result in an equation relating the
dynamics of $\phi_{\kk}(t)$ to the correlator $\langle \rho_{\kk+\qq}(t)
\rho_{-\kk}(0)\rangle$. For an equilibrium homogeneous fluid this correlator
would be zero for $\qq\neq 0$. However, for the inhomogeneous fluid this is
not the case. This makes including the influence of an external field in the MCT
a difficult problem!

\section{Conclusions}
\label{sec:conc}

In this paper we have presented a DDFT for systems of particles whose dynamics
is governed by Newton's equations. In Sec.~\ref{sec:2}, we derived
Eq.~(\ref{eq:DDFT_1}), an exact equation, which is the starting point for
our DDFT. In Sec.~\ref{sec:3} we suggested two approximations in order
to close Eq.~(\ref{eq:DDFT_1}), yielding the DDFT, Eq.~(\ref{eq:mainres}).
Then, in Sec.~\ref{sec:4} we analyzed the consequences of assuming
Eq.~(\ref{eq:mainres}), for the description of sound waves and for describing
two point correlation functions. We have shown that the DDFT incorporates the
physics described by the MCT, and the present analysis adds to the work of
Kawasaki and coworkers \cite{KawasakiPhysicaA1994, KawasakiJPCM2000,
KawasakiandKim, FuchizakiKawasakiJPCM2002, KawasakiJStatP2003, Kawasaki} in
suggesting that the DDFT provides a useful framework for studying the glass
transition. 

We believe that Eq.~(\ref{eq:DDFT_1}) needs further study. In
particular, we believe that there may be better ways to approximate the term
$B(\rr,t)$, given by Eqs.~(\ref{eq:B}) or
(\ref{eq:B_appendix}). We made the very
simple approximation $B(\rr,t)\simeq \nu \dot{\rho}(\rr,t)$. However, when the
modulations in the density profile become large we expect this approximation to
break down. Whether this can be remedied by replacing the coefficient
$\nu$ by a
function(al) of the density remains to be investigated. Having said this, the
simple approximation suggested in Sec.~\ref{sec:3} for the term $B(\rr,t)$
in Eq.~(\ref{eq:DDFT_1}), results in a dynamical theory that proves to be rather
interesting, given that it can be used as a starting point for deriving MCT.

Very recently, a DDFT (modified Cahn-Hilliard theory)
of a similar form to that in Eq.~(\ref{eq:mainres}) was
suggested in Ref.~\cite{KoideKreinRamosPLB2006}. These authors considered a
system of particles with stochastic equations of motion with correlations in the
stochastic noise. If one considers a system of Brownian particles with Langevin
equations of motion of the form
\begin{equation}
\dot{\rr}_i(t)=-\Gamma \nabla_i V(\rr^N,t)+\Gamma{\bf X}_i(t),
\label{eq:Langevin2}
\end{equation}
where $\Gamma^{-1}$ is a friction constant characterising the
one-body drag of the solvent on the colloidal particles and
${\bf X}_i(t)=(\xi_i^x(t),\xi_i^y(t),\xi_i^z(t))$ is a white noise term
with the property $\left< \xi_i^{\alpha}(t) \right> =0$ and
$\left< \xi_i^{\alpha}(t)\xi_j^{\nu}(t')\right> = 2 k_BT \delta_{ij}
\delta^{\alpha \nu} \delta(t-t')$, (where $\alpha,\nu=x,y,z$ label the
Cartesian coordinates), then one can derive a DDFT of the following
form \cite{MT,Archer7}:
\begin{equation}
\frac{\partial \rho(\rr,t)}{\partial t}=\Gamma \nabla \cdot
\left[ \rho(\rr,t) \nabla \frac{\delta F[\rho(\rr,t)]}{\delta
\rho(\rr,t)} \right].
\label{eq:MT_DDFT}
\end{equation}
However, if the noise term has memory -- i.e.~if the noise has the property
$\left< \xi_i^{\alpha}(t) \right> =0$ and
$\left< \xi_i^{\alpha}(t)\xi_j^{\nu}(t')\right> = 2 k_BT \delta_{ij}
\delta^{\alpha \nu} W(t-t')$, where $W(t-t')=\exp(-|t-t'|/\gamma)/\gamma$, then
arguing along the lines suggested in Ref.~\cite{KoideKreinRamosPLB2006},
one arrives at the following (approximate) DDFT equation:
\begin{equation}
\gamma\frac{\partial^2 \rho(\rr,t)}{\partial t^2}+
\frac{\partial \rho(\rr,t)}{\partial t}=\Gamma \nabla \cdot
\left[ \rho(\rr,t) \nabla \frac{\delta F[\rho(\rr,t)]}{\delta
\rho(\rr,t)} \right].
\label{eq:mem_DDFT}
\end{equation}
The authors of Ref.~\cite{KoideKreinRamosPLB2006} applied a linearised form of
Eq.~(\ref{eq:mem_DDFT}), together with a Ginzburg-Landau free energy functional
for $F$ to study spinodal decomposition. These authors were motivated by the
need for a causality constraint in the description of phase transitions
expected to occur in excited states of matter produced in heavy ion collisions
\cite{KoideKreinRamosPLB2006}. They found that the memory term can have a
significant effect in delaying the growth of density fluctuation in the initial
stages of spinodal decomposition. Eq.~(\ref{eq:mem_DDFT}) has the same formal
structure as Eq.~(\ref{eq:mainres}). We believe that it is worthwhile to pursue
further investigations of these equations. In this sense, the present paper is a
starting point for further developing a DDFT framework for tackling the
microscopic dynamics of fluids.

\ack
I am indebted to Bob Evans, Markus Rauscher, Andrea Gambassi, Matthias Fuchs and
Joe Brader for useful discussions and to Bob Evans for proof reading this
manuscript. I am grateful to EPSRC for support under grant number
GR/S28631/01.

\section*{Appendix}

Here we present an alternative derivation of Eq.~(\ref{eq:DDFT_1}).
The state of a system of $N$ identical particles is
specified by the set of particle position coordinates
$\rr^N=\{\rr_1,\rr_2,\cdots,\rr_N\}$ and momenta
$\pp^N=\{\pp_1,\pp_2,\cdots,\pp_N\}$.
There exists a phase space probability density
function, $f^{(N)}(\rr^N,\pp^N,t)$, which gives the probability that the system
is in a particular configuration $(\rr^N,\pp^N)$ at time $t$. The time
evolution of $f^{(N)}$ is governed by the Liouville equation \cite{HM}:
\begin{equation}
\frac{\partial f^{(N)}}{\partial t}=\{ {\cal H},f^{(N)} \},
\end{equation}
where $\{\cdot,\cdot\}$ denotes a Poisson bracket, and the Hamiltonian
\begin{equation}
{\cal H}(\rr^N,\pp^N,t)=\frac{1}{2m}\sum_{i=1}^N|\pp_i|^2+V(\rr^N,t),
\end{equation}
where $V(\rr^N,t)$, the potential energy of the
system, is given by Eq.~(\ref{eq:PE}).
Integrating over the Liouville equation, we obtain equations of motion for
reduced phase space distribution functions $f^{(n)}$, where $n<N$. One finds
that in order
to describe the dynamics of $f^{(n)}$, one must know member $f^{(n+1)}$ of
the BBGKY hierarchy \cite{HM}. The first member is the following:
\begin{eqnarray}
\left(\frac{\partial}{\partial t} + \frac{\pp_1}{m}\cdot\nabla_{\rr_1}
+ \X_1\cdot\nabla_{\pp_1} \right) f^{(1)}(\rr_1,\pp_1,t) \nonumber \\
\hspace{1cm}=-\int \dr_2\int \dpp_2 \F_{12} \cdot \nabla_{\pp_1}
f^{(2)}(\rr_1,\pp_1,\rr_2,\pp_2,t),
\label{eq:BBGKY_1}
\end{eqnarray}
where $\X_1=-\nabla_{\rr_1}V^{ext}(\rr_1,t)$ is the external force each
particle 1 and $\F_{12}=-\nabla_{\rr_1}v(\rr_1-\rr_2)$ is the pair force on
particle 1 due to particle 2.
If we integrate Eq.~(\ref{eq:BBGKY_1}) with respect to the momentum $\pp_1$,
then we obtain the continuity equation \cite{Kreuzer}:
\begin{equation}
\frac{\partial \rho(\rr_1,t)}{\partial t}+\nabla_{\rr_1} \cdot \jj=0,
\label{eq:continuity}
\end{equation}
where
\begin{equation}
\rho(\rr_1,t) =\int \dpp_1 f^{(1)}(\rr_1,\pp_1,t)
\end{equation}
is the average one--body density and
\begin{equation}
\jj(\rr_1,t) =\int \dpp_1 \frac{\pp_1}{m} f^{(1)}(\rr_1,\pp_1,t)
\end{equation}
is the current.
If we now multiply Eq.~(\ref{eq:BBGKY_1}) through by $\pp_1/m$ and then
integrate with respect to the momentum $\pp_1$, then
we obtain the following momentum balance equation \cite{Kreuzer}:
\begin{eqnarray}
\fl
\frac{\partial \jj(\rr_1,t)}{\partial t}
+ \nabla_{\rr_1} \cdot \int \dpp_1 \frac{\pp_1 \pp_1}{m^2}f^{(1)}(\rr_1,\pp_1,t)
-\frac{1}{m} \rho(\rr_1,t) \X_1 \nonumber \\
-\frac{1}{m}\int \dr_2 \F_{12} \rho^{(2)}(\rr_1,\rr_2,t)=0,
\label{eq:mom_bal_1}
\end{eqnarray}
where
\begin{equation}
\rho^{(2)}(\rr_1,\rr_2,t)
=\int \dpp_1 \int \dpp_2 f^{(2)}(\rr_1,\pp_1,\rr_2,\pp_2,t)
\end{equation}
is the two body density distribution function. Note that $\pp_1 \pp_1$ is a
tensor product (dyadic) \cite{Kreuzer}.
Taking a time derivative of Eq.~(\ref{eq:continuity}) and then
using Eq.~(\ref{eq:mom_bal_1}) to eliminate the term involving $\partial
\jj/\partial t$ we obtain the following equation:
\begin{eqnarray}
\fl
\frac{\partial^2 \rho(\rr_1,t)}{\partial t^2}=
\nabla_{\rr_1} \cdot \nabla_{\rr_1} \cdot
\int \dpp_1 \frac{\pp_1 \pp_1}{m^2} f^{(1)}(\rr_1,\pp_1,t) \nonumber \\
-\frac{1}{m}\nabla_{\rr_1} \cdot \int \dr_2 \F_{12} \rho^{(2)}(\rr_1,\rr_2,t)
-\frac{1}{m}\nabla_{\rr_1} \cdot [ \rho(\rr_1,t) \X_1].
\label{eq:mom_bal}
\end{eqnarray}
At equilibrium, from equipartition we find that the integral
$\int \dpp_1 (\pp_1 \pp_1) f^{(1)}=m k_B T
\rho(\rr_1,t) \1$, where $\1$ denotes the $3\times 3$ unit matrix.
Using this result in Eq.~(\ref{eq:mom_bal}), we obtain Eq.~(\ref{eq:DDFT_1}),
together with the following expression for $B(\rr,t)$:
\begin{eqnarray}
B(\rr,t)=-\nabla \cdot \nabla \cdot
\int \dpp\left(\frac{\pp \pp}{m^2}-\frac{k_BT}{m} \1 \right) f^{(1)}(\rr,\pp,t).
\label{eq:B_appendix}
\end{eqnarray}
It is clear from this expression, that at equilibrium $B(\rr,t)=0$. This
analysis shows that $B(\rr,t)$ is essentially the gradient of the off--diagonal
elements of the kinetic part of the pressure tensor \cite{Kreuzer}.

We also note here,
that if the potential energy $V(\rr^N,t)$ contains not just one and
two--body contributions, as given by Eq.~(\ref{eq:PE}), but also contains
three--body and higher body contributions:
\begin{equation}
\fl
V(\rr^N,t)=\sum_iV^{ext}(\rr_i,t)+\frac{1}{2}\sum_{i,j}v_2(\rr_i(t),\rr_j(t))
+\frac{1}{6}\sum_{i,j,k}v_3(\rr_i(t),\rr_j(t),\rr_k(t))+ \cdots,
\label{eq:PE_3bod}
\end{equation}
then Eq.~(\ref{eq:mom_bal}) becomes:
\begin{eqnarray}
\fl
\frac{\partial^2 \rho(\rr_1,t)}{\partial t^2}=
\nabla_{\rr_1} \cdot \nabla_{\rr_1} \cdot
\int \dpp_1 \frac{\pp_1 \pp_1}{m^2} f^{(1)}(\rr_1,\pp_1,t)
-\frac{1}{m}\nabla_{\rr_1} \cdot [ \rho(\rr_1,t) \X_1] \nonumber \\
-\frac{1}{m}\nabla_{\rr_1} \cdot \int \dr_2 \F_{12} \rho^{(2)}(\rr_1,\rr_2,t)
\nonumber \\
-\frac{1}{m}\nabla_{\rr_1} \cdot \int \dr_2 \int \dr_3 \F_{123}
\rho^{(3)}(\rr_1,\rr_2,\rr,_3,t)+\cdots,
\label{eq:mom_bal_3bod}
\end{eqnarray}
where $\F_{123}=-\nabla_{\rr_1}v_3(\rr_1,\rr_2,\rr_3)$ is the three--body force
on particle 1 due to particles 2 and 3 and where
\begin{equation}
\rho^{(3)}(\rr_1,\rr_2,\rr_3,t)
=\int \dpp_1 \int \dpp_2 \int \dpp_3 f^{(3)}(\rr_1,\pp_1,\rr_2,\pp_2,\pp_3,t)
\end{equation}
is the three body density distribution function. In the case of many body
interactions, the sum rule (\ref{eq:grad_c1}) becomes
\begin{eqnarray}
\fl
-k_BT \rho(\rr_1) \nabla c^{(1)}(\rr_1)
=\int \dr_2 \rho^{(2)}(\rr_1,\rr_2) \nabla_{\rr_1} v_2(\rr_1,\rr_2) \nonumber \\
+\int \dr_2 \int \dr_3
\rho^{(3)}(\rr_1,\rr_2,\rr_3) \nabla_{\rr_1} v_3(\rr_1,\rr_2,\rr_3)
+ \cdots \nonumber \\
=\sum_{n=2}^{\infty} \int \dr_2 \, ... \int \dr_n \rho^{(n)}(\rr^n)
\nabla_{\rr_1} v_n(\rr^n)
\label{eq:grad_c1_many}
\end{eqnarray}
Applying this sum rule as an approximation for terms on the right hand side of
Eq.~(\ref{eq:mom_bal_3bod}), together with the approximation $B(\rr,t) \simeq
\nu \dot{\rho}(\rr,t)$, we obtain Eq.~(\ref{eq:mainres}).

\end{document}